\documentclass[12pt]{article}
\usepackage{amsmath,amsfonts}
\usepackage{graphicx}
\input epsf

\makeatletter\@addtoreset{equation}{section}\makeatother

\def\del{\partial}

\def\IC{\mathbb{C}}

\def\IZ{{\mathbb{Z}}}
\def\IR{{\mathbb{R}}}
\def\IP{\mathbb{P}}

\def\CN {{\cal N}}

\def\CW {{\cal W}}

\makeatletter\@addtoreset{equation}{section}\makeatother

\newcommand{\Tr}{{\rm Tr\,}}

\renewcommand{\title}[1]{\vbox{\center\LARGE{#1}}\vspace{5mm}}
\renewcommand{\author}[1]{\vbox{\center#1}\vspace{5mm}}
\newcommand{\address}[1]{\vbox{\center\em#1}}
\newcommand{\email}[1]{\vbox{\center\tt#1}\vspace{5mm}}

\begin{document}

\unitlength = .8mm

\begin{center}
\hfill \\
\hfill \\
\vskip 1cm

\title{Quantum corrections to $\CN=2$ Chern-Simons theories with flavor and their AdS$_4$ duals}

\author{Daniel Louis Jafferis$^{1,a}$}

\address{$^1$NHETC and Department of Physics and Astronomy, Rutgers University,\\ Piscataway, NJ 08855, USA}

\email{$^a$jafferis@physics.rutgers.edu}

\end{center}

\abstract{We add fundamental flavors to $\CN=2$
Chern-Simons-matter theories living on M2 branes probing a
Calabi-Yau four-fold singularity. This is dual, in the 't Hooft
limit described by IIA string theory, to the introduction of
supersymmetric D6 branes wrapping AdS$_4$ and a 3-cycle of the
internal manifold. The resulting Chern-Simons theories remain
conformally invariant, corresponding to the fact that the D6
branes lift to pure geometry in M-theory. The determination of the
moduli space relies crucially on the 1-loop contributions to
charges and OPE's of monopole operators in these field theories.
The general picture is determined for non-chiral and chiral
flavors, and is illustrated in several examples.}

\section{Introduction}

The discovery of the eleven dimensional supergravity limit of IIA
string theory at strong coupling \cite{witten} and the web of
strong/weak dualities connecting the various known string theories
ushered in a new era of understanding of string theory. However,
the quantum mechanical description of M-theory and its brane-like
objects, the M2 and M5, has remained rather mysterious. In the
context of AdS/CFT, it was clear that the low energy conformal
field theory of $N$ M2 branes held the key to a deeper and more
detailed understanding of 2+1 dimensional conformal theories and
their M-theory gravity duals.

The lack of any tunable coupling in M-theory led naturally to the
conclusion that no Lagrangian description of the strongly coupled
conformal field theory of M2 branes existed. The origin of the M2
brane in the physics of D2 branes at strong IIA string coupling
implies that the low energy CFT is the infrared limit of the
${\cal N}=8$ 2+1 $U(N)$ Yang-Mills on $N$ D2 branes. The
Yang-Mills coupling is dimensionful in 2+1 dimensions, diverging
in the IR, corresponding to the lack of a smooth near horizon
region of the black D2 solution in IIA supergravity. Therefore
although this description is in principle complete, for example on
the lattice, it provides little insight into many aspects of the
physics relevant for AdS/CFT, such as a construction of the chiral
operators in the non-abelian theory. Furthermore, it gave little
guide to finding the theories of $N$ M2 branes at general
singularities.

The Lagrangian description of this CFT  as a Chern-Simons-matter
theory, found in \cite{ABJM} following work of Bagger and Lambert,
and Gustavsson \cite{BL, BL2, BL3, G}, exists because there are in
fact backgrounds in which the M2 branes are weakly coupled, even
in the infrared. Moreover, the appearance of Chern-Simons theories
in this context was anticipated \cite{Schwarz, GY}, as they
provide natural superconformal theories in three dimensions. The
inverse of the Chern-Simons level functions as the coupling, so
the addition of a large Chern-Simons term cuts off the running of
Yang-Mills theory, resulting in weakly coupled conformal field
theories.

In asymptotically flat space, the string coupling blows up near a
D2 brane, and, lifting to M-theory, one finds the AdS$_4 \times
S^7$ near horizon geometry of an M2 brane. Reducing to IIA on a
different circle, $U(1)_B$, which is an isometry of the full
geometry - in contrast to the usual reduction to IIA, which is
only an isometry of the asymptotic background $\IR^{9,1} \times
S^1$, not the black M2 solution - gives a background of IIA where
the D2 brane {\it does} have a smooth near horizon region, AdS$_4
\times \IC\IP^3$. Changing the background probed by the M2 branes
to a $\IZ_k \subset U(1)_B$ orbifold scales the $F_2$ flux by $k$,
and, in the large $k$ limit, results in weakly coupled IIA string
theory. It is the choice of this $U(1)_B$ isometry, required for a
Lagrangian description, which breaks the manifest supersymmetry
down to ${\cal N}=6$, which is indeed the full supersymmetry of
the orbifold AdS$_4 \times S^7/\IZ_k$ for $k > 2$.

In the large $k$ limit, the IIA description becomes weakly
coupled, $g_{IIA} \sim \frac{\lambda^{1/4}}{k}$,
 and the number of degrees of freedom of $N$ M2 branes at the
 $\IC^4/\IZ_k$ orbifold
scales as $(N k)^{3/2}/k = \frac{N^2}{(N/k)^{1/2}}$, strongly
suggesting the existence of a field theory description with $U(N)$
gauge symmetry and an 't Hooft coupling $N/k$. The near horizon
geometry in the 't Hooft limit is given by IIA on AdS$_4 \times
\IC\IP^3$ with $N$ units of $F_4$ flux (measured in terms of wedge
powers of K\"{a}hler form, $J$, on $\IC\IP^3$) and $k$ units of
$F_2$ flux in the $\IC\IP^3$, with the curvature in string units
$R_{str}^2 = 2^{5/2} \pi \sqrt{\lambda}$, to leading order.

One beautiful aspect of the AdS/CFT correspondence in this context
is that the particular features of M-theory and its reduction to
IIA string theory used above to find backgrounds with weakly
coupled M2 branes are reflected in a general property of three
dimensional gauge theories: the existence of disorder operators
carrying magnetic charge, that behave as local operators of the
CFT. They are dual to D0 branes, or, more generally, objects
carrying momentum around the M-theory circle whose specification
was required to write a Lagrangian description. Moreover, in
${\cal N}=2$ theories dual to AdS$_4\  \times $ Sasaki-Einstein
7-manifolds, these monopole operators appear in the chiral ring.
Thus the ${\cal N}=6$ $U(N)_k \times U(N)_{-k}$
Chern-Simons-matter theory and its generalizations to be discussed
below has a new regime, in addition to the 't Hooft limit which is
unsurprisingly a string theory. When $k$ is held fixed and $N$
taken to infinity, monopole operators will have low dimension,
becoming as important as mesonic operators, and the moduli space
will gain an additional large dimension, appropriate for the dual
M-theory geometry.

Many of the gauge theories that describe $N$ M2 branes probing a
Calabi-Yau singularity $X/\IZ_k$ are dimensional reductions of
${\cal N} = 1$ quiver gauge theories in 3+1 dimensions deformed by
the addition of ${\cal N}=2$ Chern-Simons terms, with the
constraint that the Chern-Simons levels sum to zero. In the
infrared, the magnetic currents $\ast\ \Tr F_i$ are conserved by
the Bianchi identity, and thus have dimension two. Therefore the
Yang-Mills term is an irrelevant operator, and can be simply
erased from the Lagrangian in the presence of a non-vanishing
Chern-Simons term, which renders the resulting action
non-singular. For general superpotentials, the chiral fields must
have R-charges that differ from free fields, thus the theory in
the infrared has strongly coupled matter sectors, interacting via
weakly coupled gauge groups, just as in the four dimensional
superconformal field theories of D3 branes at Calabi-Yau
singularities. The baryonic $U(1)_B$ isometry under which there
will be charged BPS operators is associated to the conserved
current $J = \ast\ \sum \Tr F_i$. The Chern-Simons terms imply
that the magnetic vortex is not gauge invariant, and its charge
scales with $k$. Thus the gauge invariant baryonic operators, dual
to D0 branes and their ilk, have dimension that scales with $k$,
just like the mass of the corresponding branes.

The moduli space of these theories results from the same F-term
equations as the parent 4d theory, however the D-terms are
different, associated to a sextic bosonic potential. This naively
results in a seven dimensional moduli space for a single M2 brane,
which is in fact the cone over the horizon manifold in the IIA
limit from which the weakly coupled Lagrangian emerged. At strong
coupling, that is for small $k$, the presence of light monopoles
implies that an extra dimension associated to the dual photons of
the unbroken parts of the gauge symmetry (the overall $U(1)$ in
the abelian theory) becomes large, and the full eight dimensional
moduli space appears. This perfectly matches the geometry and
$\IZ_k$ orbifold described above.

Note that even without Chern-Simons terms, there is a way to make
a three dimensional gauge field weakly coupled in the infrared.
Given a large number, $N_f$, of chiral multiplets in the
fundamental, calculations may be done perturbatively in a $1/N_f$
expansion. Moreover, the dimensions of monopole operators will
scale with $N_f$ and decouple in the limit. We will encounter
several examples of this type, in which the quiver gauge theory of
$N$ M2 branes probing a Calabi-Yau cone $X/\IZ_k$ will involve
nodes with vanishing CS level and a number of flavors that scales
with $k$. Indeed, since integrating out $N_f$ chiral fundamentals
with real (3d ${\cal N}=2$ D-term) masses shifts the Chern-Simons
level by $N_f/2$, such theories can be related by Higgsing, and
should not be fundamentally distinguished.

More generally, the $U(1)_B$ will have orbifold fixed loci in the
near horizon region so that the IIA reduction is singular, and the
dual theories likely have no simple Lagrangian description,
although there will still be weakly coupled Chern-Simons gauge
fields. My focus in the present work is the case where the
$U(1)_B$ has full fixed loci, which reduces to explicit D6 branes
in IIA string theory. They result in the addition of fundamental
flavors, and modify the moduli space quantum mechanically.

The charges of mesonic operators in ${\cal N}=2$ CSM theories are
protected, but of course, the R-symmetry in the infrared cannot
easily be guessed in the UV theory, thus their exact dimensions
are unknown without additional input. The perturbative
(non)-renormalization of these ${\cal N}=2$ is analogous to ${\cal
N}=1$ theories in four dimensions: the coefficients of terms in
the superpotential scale by the difference of the R-charge of the
operator from 2. Thus the chiral ring relations for mesonic
operators can be determined in the usual way from the
superpotential. Monopole operators, on the other hand, present a
new ingredient. Their charges can receive quantum corrections,
and, intriguingly, for the monopoles of interest to us, these
corrections are proportional to what would be anomalies of the
theory with the same quiver in four dimensions \cite{BKK}. Again,
our techniques are not powerful enough to determine  the
dimensions of these operators in general, except order by order in
perturbation theory, but I will be able to find the exact form of
their OPE's. That is sufficient to determine the chiral ring, and
thus the moduli space at the level of algebraic geometry. This
allows me to match it with the geometry, finding complete
agreement.

During the final stages of this project I became aware of a
related work \cite{BCC} that also appears on the arXiv today, in
which M2 brane worldvolume $\CN=2$ Chern-Simons-matter SCFTs with
flavor are studied from a complementary perspective.

In the next section, I describe the appearance of weakly coupled
gauge groups from M2 branes probing singularities of Calabi-Yau
4-folds, distinguishing several cases characterized by the nature
(or absence) of non-isolated fixed loci of a $U(1)_B$ isometry. In
section 3, quantum corrections to the charges of monopole
operators in ${\cal N}=2$ Chern-Simons-matter theories are
determined. The application to quivers with fundamental flavors
and their AdS$_4$ duals with explicit D6 branes is explained in
section 4. The final section illustrates these results in a few
examples.

\section{Weakly coupled gauge theories from M2 branes on CY
4-folds}

There is a natural generalization of the mechanism outlined above
to find weakly coupled gauge symmetries of M2 branes probing more
general conical singularities. In the IIA picture, these will be
the theories living on D2 branes probing seven dimensional conical
backgrounds with RR fluxes and a varying dilaton that vanishes at
the origin, such that the black D2 solution does have a smooth
near horizon region. In this paper, we content ourselves to
examples with ${\cal N}=2$ supersymmetry, as the associated
R-symmetry implies that there is a chiral ring, which controls the
geometry of the moduli space as an algebraic variety. Consider a
Calabi-Yau 4-fold, $X$, together with a $U(1)_B$ isometry that
preserves the holomorphic 4-form, ie. commutes with supersymmetry.
A toric Calabi-Yau four fold will have a $U(1)^3$ of such
isometries. Then $X /\IZ_k$ is again Calabi-Yau, with $\IZ_k
\subset U(1)_B$, and has a small circle in the large $k$ limit,
under which it is natural to reduce to IIA theory.

The worldvolume theory of M2 branes on $X$ is equivalent to that
on D2 branes in that reduction. Moreover, by taking $k$ large, the
IIA string coupling may be made arbitrarily small. Therefore the
data required to specify an $\CN=2$ Chern-Simons-matter dual CFT
is a conical Calabi-Yau 4-fold together with such a $U(1)$ action.
Different choices of the $U(1)$ isometry will give dual field
theories, generalizing 3d mirror symmetry, which in its ${\cal N}
\geq 3$ version can be geometrically interpreted as exchanging the
pair of tri-holomorphic $U(1)$ isometries of a toric
hyperK\"{a}hler 8-manifold. The matter part of the action will in
general be strongly coupled due to the superpotential, since the
matter fields must have anomalous dimensions equal to their R
charges. This is entirely analogous to a small number of D3 branes
on a Calabi-Yau 3-fold singularity when the IIB string coupling is
very weak. The essential difference is that in our case, the
inverse coupling is quantized, and, in the M-theory picture,
completely geometrized.

There are three possibilities for the $U(1)$ action on the near
horizon region: 1) it could have no fixed loci, 2) it could have a
locus of points fixed by the entire $U(1)$, or 3) there could be
loci consisting of fixed points of a discrete subgroup of the
$U(1)$. The possibilities 2) and 3) are obviously not mutually
exclusive. In this paper we are concerned with the second case.
The IIA reduction will have explicit D6 branes, but when they are
treated in the probe approximation the background IIA near horizon
geometry will be non-singular, and we expect the field theories
will have fundamental flavors.

In the simplest situation 1), the $U(1)_B$ acts transitively on
the Sasaki-Einstein horizon manifold, $S$, of $X$, thus the near
horizon geometry in the IIA reduction will be non-singular. The
dual gauge theories turn out to be described by ${\cal N}=2$
quiver Chern-Simons theories with levels that sum to zero, which I
now proceed to review.

Recall that the dimensional reduction of an ${\cal N}=1$ quiver
gauge theory with superpotential ${\cal W}$ from 3+1 dimensions to
2+1 gives an ${\cal N}=2$ Yang-Mills theory. The vector multiplet
gains an additional bosonic scalar, $\sigma$, from the component
of the gauge field along the compactified direction. The kinetic
term for the chiral multiplets includes couplings,
$-\bar\phi_i\sigma^2 \phi_i - \bar{\psi}_i \sigma \psi_i$, where
$\phi_i$ and $\psi_i$ are the bosonic and fermionic components of
the multiplet. There is the usual D term, $\bar \phi_i D \phi_i$,
inherited from four dimensions. The Yang-Mills coupling is
dimensionful, so such theories flow to strong coupling in the IR,
where emergent local operators, the monopoles discussed above, may
become important. Moreover the chiral anomalies of four
dimensional theories are not present in three dimensions, so more
general quivers are allowed, with a non-zero net number of fields
entering a node.

The IR behavior improves upon the addition of Chern-Simons terms,
$$S_{CS}^{{\cal N}=2} = \frac{k}{4\pi} \int \Tr(A \wedge dA +
\frac{2}{3} A^3 - {\bar \chi} \chi + 2 D \sigma),$$ which preserve
${\cal N}=2$ supersymmetry. The parity anomaly implies that the
Chern-Simons level must be an integer plus $1/2$ the net number of
charged Majorana fermions. Since we are interested in conformal
field theories, only massless fields will be including in the
quivers, but note that integrating out a chiral fundamental with
D-term (ie. real) mass is equivalent to shifting the Chern-Simons
level by $1/2$.

Integrating out the gauginos, $D$, and $\sigma$ fields that have
been given a mass by the Chern-Simons term, one finds the action
\begin{equation}\begin{split}
S^{{\cal N}=2} = \int  \frac{k}{4\pi} \Tr(A \wedge dA +
\frac{2}{3} A^3) +  D_{\mu} \bar\phi_i& D^{\mu} \phi_i + i
\bar\psi_i
\gamma^{\mu} D_{\mu} \psi_i \\
  - \frac{16\pi^2}{k^2}
(\bar\phi_i T^a_{R_i} \phi_i) (\bar\phi_j T^b_{R_j} \phi_j) (\bar
\phi_k T^a_{R_k} T^b_{R_k} \phi_k) &- \frac{4\pi}{k} (\bar\phi_i
T^a_{R_i} \phi_i) (\bar\psi_j T^a_{R_j} \psi_j) \\ & -
\frac{8\pi}{k} (\bar\psi_i T^a_{R_i} \phi_i) (\bar\phi_j T^a_{R_j}
\psi_j).
\end{split}
\end{equation} Note that this action has classically
marginal couplings. It has been argued that it does not
renormalize, up to shift of k, and so is a CFT. The full action
will also include the ${\cal N}=2$ superpotential terms, inherited
without change from four dimensions,  which constrain the infrared
R-charges of various mesonic operators.

The moduli space of vacua can be determined by the vanishing of
the bosonic potential, resulting in the equations
$$\del {\cal W} = 0, \ \ \ \ (k^{-1})_{i j} \mu^i T^j_{a b} Q_b = 0,$$
where $Q_b$ are the chiral fields, $\mu^i$ are the moment maps,
$i$ and $j$ are gauge groups indices, $a$ and $b$ index the chiral
fields, and $T^i_{a b}$ is the matrix given by the matter
representation of the gauge group action. The gauge fields have
been set to zero, leaving only constant gauge transformations
unfixed. Precisely which constant gauge transformations are
actually symmetries of the theory that the moduli space should be
quotiented by is a subtle question in Chern-Simons theories and
requires knowledge of the non-perturbative spectrum of magnetic
flux configurations.

The quivers of interest in this work obey the constraint $\sum_i
k_i = 0$, which implies that the moduli space of the abelian
theory is unusually large, as the moment maps may be nonzero: the
analog of the D-term equation is satisfied when $\mu_i = r k_i$,
for any $r \in \mathbb{R}$. Shifts of the dual photon of the
overall $U(1)$ under which no matter is charged act
holomorphically on the moduli space. Thus we see that simple
mesonic operators are insufficient to parameterize the entire
moduli space. Operators charged under the dual photon must carry
magnetic charge, and they can be constructed simply using the
state operator correspondence of the conformal field theory.

In particular, consider a state on the sphere with $n$ units of
magnetic flux in each gauge group, so that $\int_{S^2} F_i = 2\pi\
\textrm{diag}(n, 0, \dots)$. This corresponds to a disorder
operator which creates a vortex in flat space. The supersymmetry
variations imply that for a half BPS state, the classical vortex
solution must satisfy $$F = - \ast d \mu,$$ where the scalar in
the vector multiplet is given by $\mu = \frac{n}{2r}$ in
$\IR^{2,1}$ for a vortex located at the origin \cite{Kapustin2}.

The conformal transformation from flat space to $S^2$ implies that
a constant background of the associated adjoint scalar must be
turned on in the sphere. At the conformal fixed point, the value
of this scalar in the vector multiplet is frozen to $1/k$ times
the moment map. Therefore to construct a BPS state of the ${\cal
N}=2$ CSM theory, one needs to find an appropriate spatially
constant background of the bosonic matter fields, which must
furthermore be uncharged under the magnetic flux, otherwise those
matter fields would have to sit in angular momentum states.

In general, the Chern-Simons terms in the action imply that such a
configuration is not by itself gauge invariant, as Gauss' law is
modified, $k_i \ast F = J_i$, where $J_i$ is the matter current
coupled to the i$^{th}$ gauge group. Thus additional zero modes of
the matter fields must be excited, which are neutral under the
magnetic flux for supersymmetry to be preserved. We will abuse
notation by writing such an operator $T_{{\bf n}, M}$ as $T_{{\bf
n}} M$, where ${\bf n}$ is a weight vector and $M$ encodes the
zero modes used to form a gauge invariant operator. The mesonic
operator $M$ itself is not gauge invariant, and one must keep in
mind that $T_{{\bf n}}$ is not an honest (gauge-invariant) local
operator. However, for our purposes in finding the chiral ring, it
functions in the same way as a local field that is charged under
the gauge group. In particular, expanding an abelian Chern-Simons
term about the monopole background as
$$\sum_i \frac{k_i}{4\pi}\ 2 \int_{\mathbb{R}^1} \delta A_i\
\int_{S^2} F_i,$$ makes it clear that $T$ has charge $k_i$ under
the i$^{th}$ gauge group. More generally, $k {\bf n}_i$ is the
weight vector of the representation under the i$^{th}$ unitary
group.

For the monopoles of interest in this work, with the same flux
turned on in each gauge group, the condition $\sum_i k_i = 0$
implies that it is possible to form a gauge invariant combination
of such an operator with the bifundamental and adjoint matter
fields. Without that constraint, these monopole operators, dual to
D0 branes, would have a tadpole, corresponding to IIA string
theory with a Romans mass \cite{GT}.

In this background, the proper quantization of fermion zero modes
which are charged under the magnetic flux results in quantum
corrections to the charges and dimension of the monopole operator.
These capture 1-loop corrections to the moduli space of the
theory.

 Due to these
monopoles, only a subgroup of constant gauge transformations is
gauged on the moduli space. For our purposes, the branches of the
moduli space of interest are those in which the Hermitian moment
maps have distinct eigenvalues, naturally picking out a Cartan
subgroup. In particular, the invariance of the path integrand
requires that
$$\prod \left(e^{i\phi_i}\right)^{k_i} = 1,$$ where $e^{i\phi_i}$
are the abelian gauge transformations in the Cartan. Here we have
assumed that the Chern-Simons levels are not renormalized on the
moduli space. More generally, one can either directly compute the
metric on the moduli space at 1-loop, and fix the constant gauge
transformations by setting the dual photon to zero, or determine
the moduli space as an algebraic manifold from the chiral ring, as
we do below.

Therefore the moduli space of the abelian Chern-Simons-matter
theory, a Calabi-Yau 4-fold cone, $X$, is related to the
Calabi-Yau 3-fold cone, $Y$, that is the moduli space of the 4d
gauge theory with the same quiver by $Y = X
//U(1)_B$. Contrarywise, the eight-manifold looks like a circle bundle over a seven manifold
made out of the moduli space, $Y_{r k_i}$, of the four dimensional
abelian gauge theory resolved by FI parameters $r k_i$, warped
over the real line $\IR \ni r$ as found in \cite{JT, MS, HZ}.

A further argument for this Chern-Simons-matter description of M2
branes on $X$ was provided in \cite{aganagic}. Consider the theory
of $N$ D2 branes on $Y \times \mathbb{R}^1$, with Ramond-Ramond
2-form flux turned on. Lifting to M-theory in the infrared, this
configuration, which preserves ${\cal N}=1$ supersymmetry in 2+1
dimensions, looks like $N$ M2 branes probing a degenerate Spin(7)
8-manifold of the form $\mathbb{R}^1 \times G_2$-manifold. The
$G_2$ manifold is our $U(1)_B$ bundle over the Calabi-Yau cone
$Y$. The quiver theory describing the $N$ D2 branes before turning
on fluxes is clearly the dimensional reduction of the quiver of
$N$ D3 branes on $Y$. It was shown in \cite{aganagic} that the
usual worldvolume Chern-Simons couplings, of the form $\int F_2
\wedge S_{CS}$, on the basis branes of the quiver are turned on,
resulting in the addition of ${\cal N}=1$ Chern-Simons terms to
the 2+1 action.

To increase supersymmetry to ${\cal N}=2$, on must further add the
couplings $\frac{k_i}{2\pi} \int D_i \sigma_i$. Likewise, in the
geometry, warping $Y$ over $r \in \mathbb{R}^1$, with the Kahler
parameters of $Y$ given by the moment maps $\mu_i = k_i r$,
results in the Calabi-Yau 4-fold $X$. Therefore it is natural to
conclude that these operations are equivalent, essentially
deriving the quiver CSM description of the M2 theory. Note that
this assumed the absence of explicit D6 branes in the IIA
reduction on $U(1)_B$, furthermore, if the IIA geometry has
non-isolated singularities more general ``fractional'' RR fluxes
are possible that cannot be including by simply adding
Chern-Simons terms.

Generic $U(1)_B$ isometries will have orbifold fixed loci on $S$,
and the near horizon geometry is singular in IIA string theory.
The theory of D3 branes probing $X//U(1)_B$ still makes sense, but
the moduli space of any Chern-Simons-matter theory based on it
will have non-isolated singularities in its eight dimensional
abelian moduli space. The presence of the singularities in $M_6 =
X/U(1)_B$ allows ``fractional'' Ramond-Ramond fluxes to be turned
on, so that the total space of the circle bundle is smooth. It
appears that the associated description of the M2 brane theory
does involve weakly coupled gauge fields in the large $k$ limit,
however they couple strongly interacting matter sectors with no
Lagrangian description.

A simple example of this is the orbifold $\IC^4/\IZ_q$, where
$\IZ_q$ acts via multiplication by $(\zeta, \zeta^{-1}, \zeta^p,
\zeta^{-p})$, for $\zeta$ a $q^{th}$ root of unity and $p$, $q$
relatively prime. This cone with an isolated singularity preserves
${\cal N}=4$ supersymmetry, and can be engineered by T-dualizing
and lifting to M-theory a configuration of NS5 brane and $(p,q)$
fivebrane intersecting D3 branes wrapping a circle \cite{GGPT}.
The recent work of \cite{GW} gives a prescription for finding the
conformal field theory in such cases with $\CN=4$ supersymmetry
involving multiple D3 branes stretched between $(p,q)$ fivebranes,
generalizing the results of \cite{KOO, BHKK} on the
Yang-Mills-Chern-Simons theory that arises on D3 branes stretched
between $(1,k_i)$ fivebranes.

The essential idea can be illustrated most simply if $q = m p  +
1$ for some $m$. First, note that application of the $SL(2, \IZ)$
transformation $T^m S T^p$ to an NS5 brane results in a $(p, p m +
1)$ fivebrane. Thus we can imagine that the $N$ D3 branes wrap a
circle and intersect a $(1,m)$ 5 brane and a $(1,p)$ 5 brane, with
S-duality transformations applied between each. The action of
S-duality was determined beautifully in \cite{GW}, in terms of
coupling to a three dimensional conformal field theory,
$T(SU(N))$, with a $U(N) \times U(N)$ flavor symmetry.

The self mirror theory $T(SU(N))$ was defined as the infrared
limit of an $\CN=4$ $U(1) \times U(2) \times ... \times U(N-1)$
Yang-Mills theory with a bifundamental hypermutliplet between each
node and $N$ fundamental hypers on the final node. The latter
carry an obvious $SU(N)$ flavor symmetry, while the other $U(N)$
symmetry is its mirror, emerging at strong coupling on the Coulomb
branch. Thus the theory describing $N$ M2 branes probing this
singularity is given by the quiver shown in figure 1, where the
lines represent hypermultiplets. The Chern-Simons levels result
from the $T$ transformations, as explained in \cite{KOO, BHKK}. As
expected it has a pair of weakly coupled Chern-Simons gauge fields
in the large $q = m p +1$ limit, but the matter sector appears to
have no Lagrangian description.

\begin{figure}
\centering
\includegraphics[height=2in.]{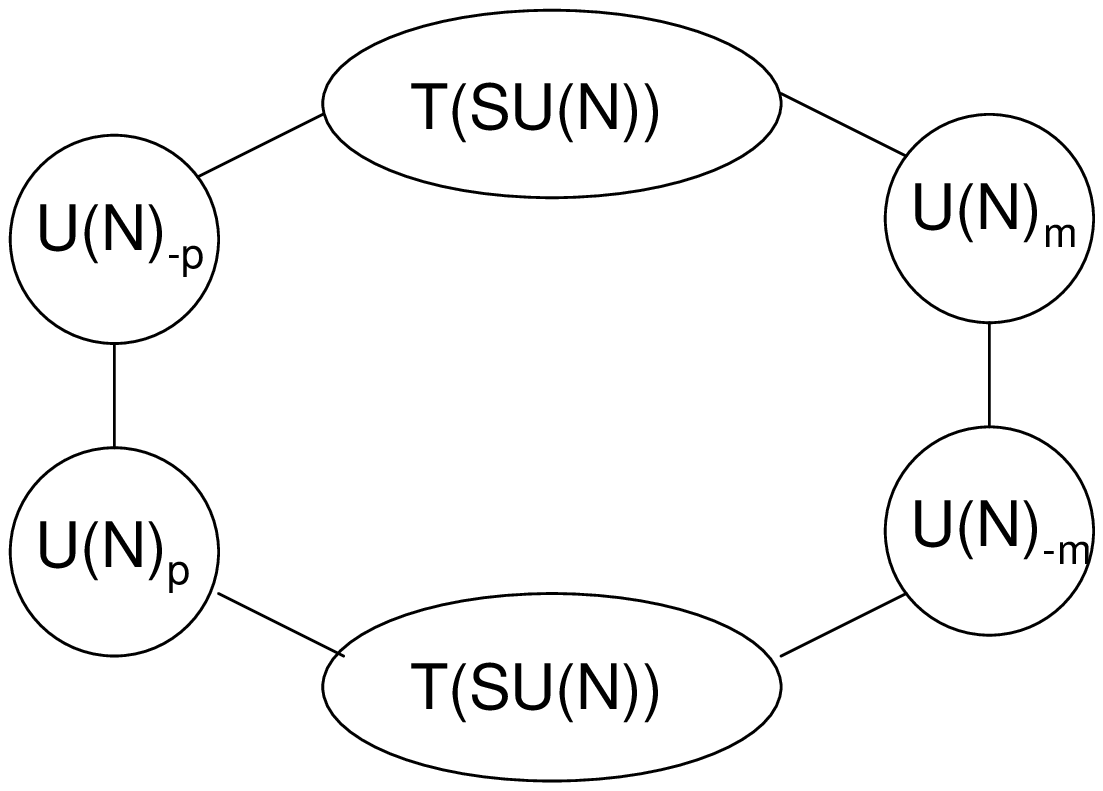}
\caption{Generalized quiver dual to $\CN=4$ AdS$_4 \times
S^7/\IZ_{p m +1}$.}
\end{figure}

In this work I will examine an intermediate situation, in which
the entire $U(1)_B$ shrinks on 3-cycles in $S$. Then the reduction
to IIA is still a nonsingular manifold, $M_6$,  however explicit
D6 branes will wrap those 3-cycles, in the probe limit. Adding
multiple coincident D6 branes will lead to an orbifold singularity
in the M-theory lift. I will show that the worldvolume theory of
D2 branes in such a background is again related to the dimensional
reduction of the theory of D3 branes probing $Y = X//U(1)_B$,
which is now a conical Calabi-Yau 3-fold with an isolated
singularity, but with the inclusion of fundamental flavors arising
from the 2-6 strings in addition to ${\cal N}=2$ Chern-Simons
terms.

\section{Quantum corrections to Chern-Simons-matter theories}

The renormalization group properties of $\CN=2$ gauge theories in
three dimensions are analogous to $\CN=1$ theories in four
dimensions. The superpotential only renormalizes through a
logarithmic scaling of coefficients by an amount proportional to
the difference of their R-charge from 2. At the conformal fixed
point, this provides a constraint on the infrared R-charges of the
fields. It is possible that a coefficient in the superpotential
could run to zero at low energies, resulting in a flow to a
seemingly completely different geometry, see
\cite{martelli-sparks} for an example. Another subtlety is that
certain Chern-Simons-matter theories break supersymmetry, and  UV
YM-CSM of that form may either have no supersymmetric vacuum, or
undergo a cascade \cite{ABJ, Aharonyetal}. In the examples
considered in this paper, such issues will not arise.

The OPE's of mesonic operators are thus uncorrected, and the
F-term equations are simply $\del \CW = 0$. As explained, the full
chiral ring is much larger, containing BPS disorder operators that
behave as local fields in the infrared. It is more difficult to
determine their OPE's directly, but the chiral ring relations are
strongly constrained by consistency of various charges on both
sides of the equation. The R-charge, which agrees with the
conformal dimension in the IR, will play a particularly important
role. Moreover, the conjecture we will make below agrees perfectly
with the geometric picture of the M-theory lift of the flavor D6
branes.

\subsection{Monopole operators in $\CN=2$ CSM without flavor}

We will regulate the CSM theory by adding a Yang-Mills term, which
is irrelevant at low energies. All calculations will be performed
in the UV theory, where the gauge fields and matter fields become
free, but the results are protected by ${\cal N}=2$ supersymmetry.
Since the theory is a CFT in the IR, the operator dimensions are
given by the infrared R-charges, which differ from those in UV
theory.

The 1-loop correction to the charge of a monopole operator under a
given $U(1)$, which can be a flavor symmetry, gauge symmetry, or
R-symmetry, can be determined using the state operator
correspondence. The monopole operator corresponds to a particular
state on $S^2$ via the radial quantization of the theory, in which
$\int_{S^2} F$ is the monopole charge. As was shown in
\cite{Kapustin1, Kapustin2, Borokhov}, the spectrum of
electrically charged fermions is altered in such a background,
resulting in the possibility of a quantum correction to the $U(1)$
charge of the vacuum in this sector.

Actually performing such a calculation in the conformal theory can
be rather involved, however supersymmetry guarantees that the same
result will be obtained by determining the fermion zero mode
contribution to the charge of the state on the sphere in the UV
theory. Following \cite{Kapustin1, Kapustin2}, the spectrum of
fermion zero modes charged under the magnetic $U(1)$ is then given
by $E_p = \textrm{sign}(p) (\frac{1}{2} |n q_e| + p)$, for $p = 1,
2, ...$, and $E_0 = \frac{1}{2} |n q_e|$, where $n$ is the number
of units of magnetic flux, and $q_e$ is the charge of the fermion
under that $U(1)$. The bosonic spectrum is entirely symmetric. The
asymmetry in the fermion spectrum means that the vacuum will carry
the charges of the unpaired mode.

Therefore the quantum correction to the charge of the monopole is
given by
$$-\frac{1}{2} \sum_{fermions} |q_e| Q,$$ where $Q$
is the charge of the fermion under the $U(1)$ of interest. The
effect of this on the chiral ring translates into a 1-loop
contribution to the metric on the moduli space, obtained by
integrating out chiral and vector multiplets that become massive
at generic points on the moduli space. If these fields are coupled
to a gauge field whose dual photon is part of the moduli, which is
the case when the associated monopole operators are BPS and
enlarge the chiral ring, then the metric is modified in a way
entirely analogous to the Coulomb branch of 3d Yang-Mills theories
\cite{jafferis-yin}. In fact, the formula above can be applied in
that case as well.

The monopole that is relevant for the geometric branch of the
Chern-Simons-matter theories describing M2 branes on 4-folds has
magnetic flux turned on in the $\textrm{diag}(1, 0, 0, ...)$ of
each gauge group. Suppose that we knew the $U(1)_R$ of the
associated 4d gauge theory. The correct IR R-symmetry of the CSM
theory may consist of this charge plus some combination of flavor
$U(1)$ charges.

Naively it appears that no useful information can be extracted
from this method, since one would have to guess the exact IR
R-symmetry in the UV theory. However for the special class of
monopoles considered here, we will show that they are neutral
under all flavor symmetries of the UV theory. Thus if we assume
that the IR R-symmetry is not an accidental symmetry of the low
energy theory, the dimension of the monopole operator determined
by our calculation does not depend on the mixing of the R-charge
with flavor symmetries, and we find a unique answer for physical,
gauge invariant operators (up to a contribution that depends only
on the monopole charge itself) which, moreover, passes a variety
of consistency checks.

In theories with at least $\CN=3$ supersymmetry, the dimensions of
the matter fields are not renormalized, due to the existence of a
non-abelian R-symmetry. The quantum correction to the dimension of
the monopoles in that case can be determined exactly \cite{GW,
GJ}, and is given by $$ \frac{1}{2} (\sum_{i\in hyper} - \sum_{i
\in vector}) |q_i| ,$$ where the $q_i$ are the gauge charges of
fermions in either hypermultiplets or vectormultiplets under the
magnetic $U(1)$.

In a non-chiral theory, the R-symmetry cannot mix with the
emergent $U(1)_B$ in the infrared, since there is a symmetry that
relates the monopole to the anti-monopole, while exchanging the
roles of each chiral field with its conjugate. In chiral theories,
such mixing is possible, but conservation of monopole charge means
that this will not impede our ability to constrain the OPE's by
matching dimensions of operators.

Suppose the gauge groups are all of equal rank. Then the quantum
correction to the UV R-charge of the monopole is given by
$$-\frac{2(N_c-1)}{2} \sum_{fermions} \textrm{R-charge},$$ where the sum is over the chiral and vector multiplets.
This quantity is proportional to the conformal anomaly of the 4d
theory, as seen in \cite{BKK}, which vanishes by consistency.
Suppose we consider some flavor symmetry of the theory, acting on
the chiral multiplets. Then the induced charge of the monopole
will be given by
$$-\frac{2(N_c-1)}{2} \sum_{chirals} \textrm{flavor charge} = 0,$$
for toric quivers. This is because the superpotential must be
invariant, and each chiral field appears exactly once in two terms
of the superpotential. Thus the sum of the flavor charge over all
the chiral fields must be zero, and the vector multiplets are
neutral under the symmetry. In generic four dimensional theories,
if this sum was non-vanishing, the flavor symmetry would be
anomalous.

This shows that whatever the correct combination of the UV
R-charge and flavor charges that gives the R-symmetry in the IR
 is, this monopole operator has no quantum correction to its
dimension, with one exception. The R-current of the conformal
field theory might include contributions from the emergent
baryonic symmetry itself. Regardless, the constraints on OPE's
from matching dimensions of both sides that I will shortly use to
determine the chiral ring are unchanged, since the ring is graded
by the monopole charge. In particular, the OPE of $T$ and $\tilde
T$ in terms of mesonic operators will have cancelling
contributions from the monopole charge to the R-charge of $T$ and
$\tilde T$.

Naively, there is an ambiguity in our prescription for determining
the quantum correction to the flavor and R-charges of monopole
operators, since the charges of the matter fields are only defined
up to gauge equivalence. This does not arise in the theories
without fundamental flavors (or even with non-chiral flavors) when
all of the ranks are equal, since the sum over all chiral
multiplets of any gauge charge is zero. When the ranks are
unequal, however, this would appear to present a paradox, which is
resolved as follows.

Consider a quiver of unitary gauge groups, $U(N_i)$, with $m_{i
j}$ chiral fields in the $(\bar{N_i}, N_j)$ representation of
dimension $R_{i j}$ - these should be understood as the exact
infrared R-charges. Then the quantum correction to the dimension
of the monopole operator, $T$, is given by $$-\frac{1}{2} \sum_i 2
(N_i - 1) - \frac{1}{2} \sum_{i, j} m_{i j} (R_{i j} - 1)(N_i+N_j
-2),$$ since the fermions in the vector multiplet have dimension
$1/2$, and in the chirals, the fermion has R-charge one less than
the bosonic component.

Suppose the assignment of R-charges to the matter fields is
shifted by some multiples of the gauge charges, $R_{i j}' = R_{i
j} + \theta_i - \theta_j$. This should change nothing about the
physics, but it seems to alter the quantum correction to the
monopole dimension. However, in this situation with unequal ranks,
the number of ingoing and outgoing chiral fields from each gauge
group will not be equal, and the monopole will receive 1-loop
corrections to its gauge charges. This is closely related to the
fact that in the monopole background, the VEV of  the associated
moment map is nonzero, giving these chirals charged under the
magnetic flux a mass. Integrating them out shifts the Chern-Simons
levels, and thus the charges of the monopole under the gauge
group.

More systemically, the charge of the monopole under the overall
$U(1)$ of the i$^{th}$ gauge group is given by
$$Q_i = \frac{1}{2} \sum_j m_{i j} (N_i + N_j - 2) - \frac{1}{2}
\sum_{\ell} m_{\ell i} (N_i + N_j - 2) \pm k_i,$$ for $T$ and
$\tilde T$ respectively. It would be interesting to determine the
precise representation of $U(N_i)$ the monopole operator lived in,
but that full non-abelian calculation will be left for future
work. Given the charges under the overall $U(1)$'s, a gauge
invariant combination must involve $n_{i j}$ zero modes of the
$(\bar{N_i}, N_j)$ chiral fields satisfying the condition
$$Q_i - \sum_j n_{i j} + \sum_{\ell} n_{\ell i} = 0.$$

Given the new assignment of R-charges, shifted by a gauge
transformation, the R-charge of the mesonic operator, $M$, made of
the $n_{i j}$ matter fields shifts as $$R'(M) = R(M) + \sum_{i, j}
n_{i j} (\theta_i - \theta_j),$$ while the quantum contribution to
the R-charge of the monopole becomes $$R'(T) = R(T) - \frac{1}{2}
\sum_{i, j} m_{i j} (\theta_i - \theta_j) (N_i + N_j - 2).$$

Therefore the dimension of the gauge invariant combination shifts
by precisely $\sum \theta_i k_i$. This is not zero in general, but
corresponds precisely to a mixing of the UV R-symmetry with the
emergent $U(1)_B$. That is, only the kernel of $\beta$ defined
above is gauged, and shifting the action of the R-symmetry by
transformations in $\ker \beta$ does not charge the above
calculation. So everything is consistent, but the exact dimension
of the monopole cannot be determined in terms of the dimensions of
the matter fields. As explained before, however, the grading of
the chiral ring by monopole charge implies that the OPE's of gauge
invariant operators are completely determined.

\section{D6 branes and fundamental flavors}

Consider a quiver CSM theory dual to AdS$_4 \times$SE$_7$ with a
transitive $U(1)_B$ action. It is given by the 2+1 dimensional
reduction of the 3+1 quiver gauge theory describing D3 branes on
the tip of the Calabi-Yau 3-fold cone $Y = X//U(1)_B$, together
with $\CN=2$ Chern-Simons terms, with levels summing to zero.

Introduce an $\CN=2$ fundamental, $q$, and anti-fundamental,
$\tilde q$ on the node with gauge group $U(N_i)$, and add the
$\CN=2$ superpotential ${\cal W} = q f p + m (qp)^2$, where $f$ is
a a mesonic operator of the original theory in the adjoint
representation of $U(N_i)$. At this stage $m$ might be zero; its
role in the Higgs branch of the moduli space will be explained
below. These fundamentals must correspond to the 2-6 strings that
become massless when the D2 brane sits on the D6 brane - in flat
space such a system is T-dual to the D4-D0 ADHM theory.

The strategy will be to determine the quantum corrected chiral
ring, and thus the back-reacted Calabi-Yau 4-fold at the level of
algebraic geometry. A non-trivial check will be that this geometry
is indeed the lift of the configuration of D6 branes. Using the
calculations of the charges of monopole operators in the previous
section, I will be able to constraint the OPE's. Working in the
abelian theory for simplicity, the chiral ring will be determined.

I will also consider examples with chiral flavors. In order for
the renormalized Chern-Simons levels to sum to zero on the moduli
space, so that there is an M-theory description, the total number
of fundamentals and anti-fundamentals must be equal, but they need
not connect to the same node in the quiver. The basic example is a
fundamental $q$ of $U(N_i)$ and an anti-fundamental $\tilde Q$ of
$U(N_j)$, together with a superpotential ${\cal W} = q f \tilde
Q$, where now $f$ is a mesonic operator in the $(\bar{N_i}, N_j)$
representation.

Note that in the non-chiral case, the pair of flavors can be
placed at any of the nodes in the quiver involved in $f$, while in
the chiral case only one choice is allowed. This corresponds to
the fact that the horizon manifold of a degree zero conical
4-cycle, cut out by a gauge invariant $f = 0$  in the conical
Calabi-Yau 3-fold, $Y$, may have a discrete $\pi_1$. Different
choices for the discrete Wilson line on the D6 brane determine the
location of the flavor pair. Such Wilson lines for the D6
worldvolume gauge field lift to a topologically non-trivial flat
C-field in M-theory, combining with the torsion fluxes arising
when the ranks of the gauge groups are unequal. The horizon
manifolds of higher degree conical 4-cycles, as in the chiral
case, are simply connected.

\subsection{Higgs branch moduli space}

If the number of flavors is greater than 1, there will be a Higgs
branch, on which the flavors get VEVs. This corresponds to D2
branes dissolving into instantons of the $SU(N_f)$ worldvolume
theory of the D6 branes. In the M-theory description, there is an
$A_{N_f-1}$ orbifold singularity which can support fractional M2
branes, into which the ordinary M2 branes split into on the Higgs
branch.

The fundamentals become massless along the locus in the moduli
space where the ``geometric'' branch is connected to the Higgs
branch. From the bosonic potential resulting from the Chern-Simons
type D-terms, there is a contribution to the mass of the
fundamentals given by
$$\frac{1}{k^2} \mu_i^2,$$ where $\mu_i$ is the moment map on that
node. The superpotential adds a mass of the form $|f|^2.$

Therefore the fundamentals become massless exactly when $$f = 0
\textrm{ and }\mu_i = 0.$$ This is the location of the D6 brane in
the conical seven dimensional IIA geometry, $\textrm{Cone}(M_6) =
X/U(1)_B$. ${\cal N}=2$ supersymmetry is preserved if the 4-cyles
wrapped by the D6 branes, which sit in the Calabi-Yau 3-fold $Y$,
given that the real moment maps are set to zero, are holomorphic.
The presence of the massless 2-6 strings indicates that the 1-loop
correction to the metric on the geometric branch of the moduli
space actually changes the topology near that locus. Moreover, the
Higgs and geometric branches join, in a singular manner, along
that intersection.

The Higgs branch is determined by the equations $\del {\cal W} =
0$, for the full superpotential including ${\cal W}_{fl}$,
together with the K\"{a}hler quotient by all of the gauge groups,
that is, it is the moduli space of the same quiver interpreted as
a four dimensional gauge theory. Given the form of the
superpotential, when the fundamentals have VEVs, one must have
$f=0$, thus the Higgs branch looks like the moduli space of $N_c$
instantons of rank $N_f$ on the complex surface determined by
those equations. That is exactly what is expected from $N_c$ D2
branes dissolved into $N_f$ D6 branes wrapping such a two complex
dimensional surface.

When the Chern-Simons theory with flavors has $\CN=3$
supersymmetry, the wrapped cycles are hyperK\"{a}hler, and no
Higgs branch exists in the abelian D6 theory (ie. when $N_f = 1$)
\cite{GJ}. In $\CN=2$ language, this is due to the quartic term,
$m (q \tilde q)^2$, in the superpotential. Typically in $\CN=2$
theories, this term is not present, since together with $q f
\tilde q$ it would constrain the operator $f$ to have dimension 1
in the IR, which, in general, would be inconsistent with the
original superpotential. Thus if $f$ had too small a dimension,
$m$ would run to zero in the infrared, while if $f$ had too large
a dimension, the fundamentals would decouple.

However, without this term in the superpotential, is it sometimes
possible that there is a Higgs branch in the field theory with
even a single non-chiral flavor
pair. 
This should only occur when the dual Sasaki-Einstein manifold is
singular, allowing a single M2 brane to fractionate.

 To satisfy the
D-term equations with only a single flavor, one must have $|q|^2 =
|\tilde q|^2$. The F-term equations on such a Higgs branch can be
solved when $f=0$, $\del f = 0$, and $\del {\cal W}_{unflavored} =
0$. In principle, there could be other solutions, of the form
$f=0$ and $\del {\cal W}_{unflavored} = - (\del f) q \tilde q$,
however it is usually impossible to satisfy such an equation
without both sides separately vanishing\footnote{There are $g+2$
relations among the F-term equations of an unflavored abelian CSM
theory with a Calabi-Yau 4-fold `moduli space, where $g$ is the
number of gauge groups. Thus the flatness of ${\cal W}$ in the
directions that do not appear in $f$ typically set all $\del \CW =
0$. It would be interesting to understand the situations when this
is not the case.}.

The equations $f=0$ and $\del f = 0$ are precisely the conditions
that the 4-cycle wrapped by the D6 branes has a singularity. Along
that locus there is a non-abelian flavor group, and in the
M-theory lift, the near horizon geometry is singular.


The Higgs branch of a theory with $N_f$ chiral flavors $q$ and
$\tilde Q$ is very similar. The fundamentals become massless when
$f = 0$ and $\mu_i=0$, which implies that $\mu_j = 0$ given the
structure of the Chern-Simons D-terms.
The moment maps of the abelian theory {\it excluding} the
contributions from the fundamentals, labelled $\mu_{\ell}'$, are
set to zero on this branch of the moduli space, with the exception
of $\mu_i' = - \mu_j' = \sum |q_a|^2 = \sum |\tilde Q_a|^2$. This
branch of the moduli space can be described as a space, $M$,
fibered over the VEV's of the fundamentals $q_a$. The phases of
$\tilde Q_a$ can be rotated relative to the phases of  $q_a$ by a
gauge transformation that acts trivially on the other fields. The
space $M$ is parameterized by the fields of the quiver excluding
the fundamentals, at the values of the FI parameters given by the
above $\mu'$ together with the equation $f = 0$. The fibration is
non-trivial since the gauge groups act both on the fundamentals
and the other fields. The fiber above $q_a = 0$ is precisely the
4-cycle in the IIA cone wrapped by the D6 brane. If there is only
a single pair of chiral flavors, then, as in the non-chiral case,
in general the Higgs branch requires that $f=0$ and $\del f = 0$,
which can only be satisfied if the D6 branes are wrapping a
singular 4-cycle.

\subsection{Monopole dimensions and OPE with flavor}

Suppose we have added $N_f$ fundamental flavors. The flavor
symmetries of the original quiver act trivially on them, and their
own flavor symmetries are always nonabelian groups, which cannot
mix with the R-symmetry. Thus we are justified in computing the
quantum correction to the naive UV R-charge of our monopole. This
results in $$-\frac{2 N_f}{2} (d_{fund} - 1) = \frac{N_f}{2}
\textrm{dimension}(f),$$ since the total dimension of the
superpotential $q f p$ must be 2.

The OPE of two monopole operators carrying opposite monopole
charge can be computed using a cylinder diagram, on $S^2 \times
I$, with magnetic flux on the $S^2$ \cite{Kapustin1, Kapustin2}.
In the calculation of the OPE for $T$ and $\tilde T$, the mesonic
chiral operators with a nonvanishing 1-point function in this
background can appear on the right hand of the product. Given the
charges of the monopoles calculated above, the possibilities are
extremely limited. This leads us to conjecture the following
simple form of this monopole/anti-monopole OPE,
$$T \tilde T \sim f^{N_f}.$$ Moreover, in the abelian theory,
the chiral ring products of monopoles with nonzero total $U(1)_B$
charge do not give any new relations \cite{GJ}, they merely relate
the monopoles with $n$ units of flux to powers of those with 1
unit of flux.

The geometry associated to this quantum corrected chiral ring,
$X$, can be expressed in a simple way in terms of the classical
moduli space, $X_c$. Recall that it is defined by the equations
$\del W = 0$, together with the K\"{a}hler quotient by the kernel
of the map $\beta: U(1)^{N_c} \rightarrow U(1)$, which sends
$\{e^{i\phi_i}\} \mapsto e^{i \sum k_i \phi_i}$. The baryonic
symmetry of $X_c$ is the quotient $U(1)^N/\ker(\beta)$. Including
the monopole operators $t$ and $\tilde t$ in the chiral ring, one
should perform the full $U(1)^N$ K\"{a}hler quotient, but note
that the monopoles are invariant under $\ker \beta$, since they
have charges precisely $k_i$ ($-k_i$ for $\tilde t$).

Therefore the quantum corrected moduli space is
$$X = (X_c \times \mathbb{C}^2)//U(1), \ t \tilde t = f^{N_f},$$
where the $U(1)$ acts as $U(1)_B$ on $X_c$ and with weights $\pm
1$ on $\mathbb{C}^2$, beautifully matching with the M-theory lift
of the D6 brane configuration. The baryonic isometry of this
quantum corrected moduli space is just monopole charge, that is is
acts as $(t, \tilde t) \mapsto (e^{i\phi} t, e^{-i\phi} \tilde
t)$. The fixed points of this $U(1)_B$ isometry, which is the
rotation of the M-theory circle, is exactly the locus where the D6
branes were wrapped, $f=0$ in $X//U(1)_B = X_c//U(1)_B$, as
characterized by the vanishing mass of the fundamental 2-6
strings.

We now turn to the case of chiral flavors, still requiring that
the total number of incoming and outgoing arrows for the entire
quiver are equal, so that the quantum corrected Chern-Simons
levels sum to zero on the moduli space and the dual geometry can
be lifted to M-theory. Here the fact that there is a 1-loop
correction to the gauge charge of the monopole operators is
crucial even in the abelian theory. These quivers cannot exist in
four dimensions, as the gauge groups would have chiral anomalies.
However, they can be regarded as dimensional reductions of
consistent 4d Yang-Mills theories with certain real masses that
only exist in three dimensions, turned on.

The calculation of the quantum correction to the dimension of the
monopole operators works the same as in the chiral case, but the
presence of the unpaired chiral fundamentals implies that the
charge under the $\ell^{th}$ abelian gauge group is $\pm k_{\ell}
+\frac{N_f}{2} {\delta_{i \ell}} - \frac{N_f}{2} {\delta_{j
\ell}}$ for $T$ and $\tilde T$ respectively. Therefore the form of
the OPE is identical to the case of non-chiral flavors, $$T \tilde
T \sim f^{N_f},$$ where now both sides are in $N_f$ times the
bifundamental representation of $U(1)_i \times U(1)_j$.

The complete chiral ring is given by the chiral multiplets
together with $t$ and $\tilde t$, with the relations $\del \CW =
0$, the above relation on $t\tilde t$, and a K\"{a}hler quotient
by the full $U(1)^{N_c}$ gauge group, acting on $t$ and $\tilde t$
with the charges above.

\section{Examples / IIB brane constructions}

Given the machinery just developed, it is extremely
straightforward to apply it to many examples, both in the
direction of determining the moduli space of a given flavored CSM
theory, and in finding all of the dual Lagrangian descriptions of
M2 branes probing a specified Calabi-Yau 4-fold cone. As
explained, most supersymmetry preserving $U(1)$ isometries will
have orbifold fixed points in the near horizon regime, and are not
suitable for our discussion. Roughly speaking, abelian isometries
acting with weight $\pm 1$ lead to smooth IIA reductions, and
those with some weights equal to 0 have D6 branes; higher weights
lead to singularities in the IIA description.

\subsection{$\CN=2$ embedding of D6 branes in $\IC\IP^3$}

Recall that the dual SCFT to IIA theory on AdS$_4 \times \IC\IP^3$
is the $U(N)_k \times U(N)_{-k}$ gauge theory with $\CN=6$
supersymmetry and a pair of bifundamental hypermultiplets
\cite{ABJM}. This theory shares its quiver diagram with the four
dimensional gauge theory of $N$ D3 branes probing the conifold.
This is no surprise given their derivations in terms of similar
fivebrane setups.

Applying T-duality to $N$ D4 branes wrapping a circle and
intersecting a pair of NS5 branes at angles preserving $\CN=1$
supersymmetry in 3+1 dimensions and zooming in results in D3
branes at the conifold singularity. Likewise, T-dualizing and
lifting to M-theory a configuration of $N$ D3 branes on a circle
intersecting an NS5 brane and $(1,k)$ 5 brane at $\CN=3$ angles
gives $N$ M2 branes probing $\IC^4/\IZ_k$. This also fits in the
general picture of $\CN=2$ Chern-Simons-matter theories, since
$\IC^4// U(1)_B$ is precisely the conifold, where $\IZ_k \subset
U(1)_B$.

The addition of D6 branes to AdS$_4 \times \IC\IP^3$ preserving
$\CN=3$ supersymmetry has been investigated in \cite{GJ, Kirsch,
WeiLi, Fujita}; here we consider a different embedding that
preserves only $\CN=2$, that was analyzed in \cite{CP, AEMOW}.
These are 2+1 analogs of the theories studied in \cite{ouyang},
and have a similar fivebrane engineering construction. The
M-theory lift can be easily determined from the general results of
the previous sections.

Consider the embedding $A_1 B_1 = 0$ in $\IC\IP^3$ with projective
coordinates $A_1, A_2,$ $B_1^*, B_2^*$. This has two branches that
intersect over an $S^2$, in contrast to the $\CN=3$ configuration
of a D6 brane wrapping a single $\IR\IP^3$, $A_1 B_1 + A_2 B_2 =
0$. It can be deformed into a smooth 3-cycle inside $\IC\IP^3$,
however it would not then be conical, and conformal invariance
would be broken. In fact, there are two types of deformations of
this embedding, one holomorphic and connected to the $\CN=3$
embedding, and the other a kind of blow-up, which will be
discussed in the final subsection of this paper.

A non-chiral pair of fundamental chiral multiplets will be
introduced into the $\CN=6$ quiver, with a superpotential $\CW = q
A_1 B_1 \tilde q$. This quiver theory, shown in figure 2, is
exactly the 2+1 dimensional reduction of the four dimensional
field theory found by \cite{ouyang} to describe a D7 brane in
AdS$_5 \times T^{1,1}$, wrapping the cycle $A_1 B_1 = 0$ in the
conifold, together with $\CN=2$ Chern-Simons terms. Moreover, the
K\"{a}hler quotient of the M-theory geometry by $U(1)_B$ is the
conifold, with D6 branes wrapping the same cycle.
 Applying the general calculation of the chiral ring implies that
this theory describes M2 branes in a Calabi-Yau 4-fold cone cut
out by the equation $t \tilde{t} = a_1 b_1$ in $\IC^6//U(1)$,
where the group acts via
$$(a_1, a_2, b_1, b_2, t, \tilde t) \mapsto (\lambda a_1, \lambda
a_2, \lambda^{-1} b_1, \lambda^{-1} b_2, \lambda^k t, \lambda^{-k}
\tilde t).$$ This manifold is toric, since the equation can be
rewritten as a Kahler quotient. In particular, it is given by
$\IC^6//U(1)^2$, acting with weights $$\begin{array}{cccccc} \ 1 \
&-1\ &1\ \ &-1\ &\ 0\ \ &\ 0\ \ \\ k+1&-k\ &0\ \ &-1\ &1\ &-1\
\end{array}.$$

\begin{figure}
\centering $\begin{array}{ccc} \includegraphics[height=.7in.,
width=2.2in.]{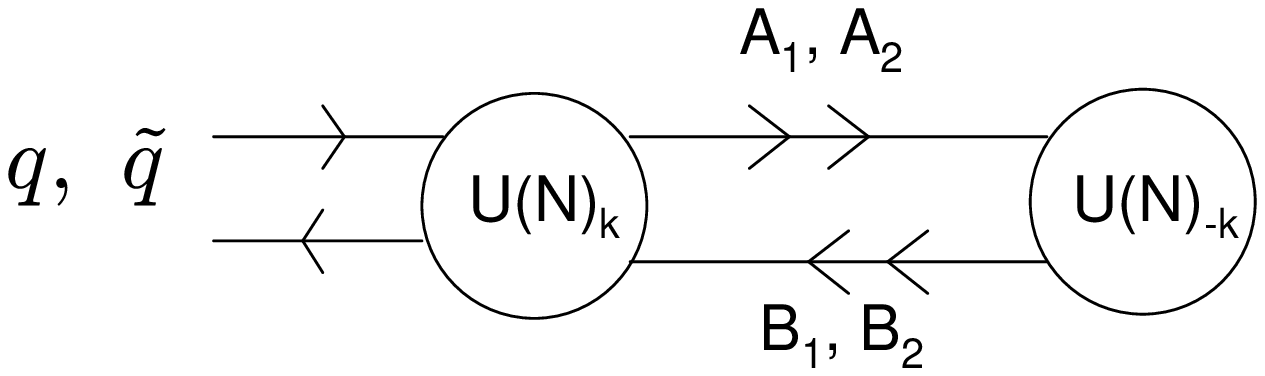} & \hspace{.3in.} &
\includegraphics[height=.6in., width=2in.]{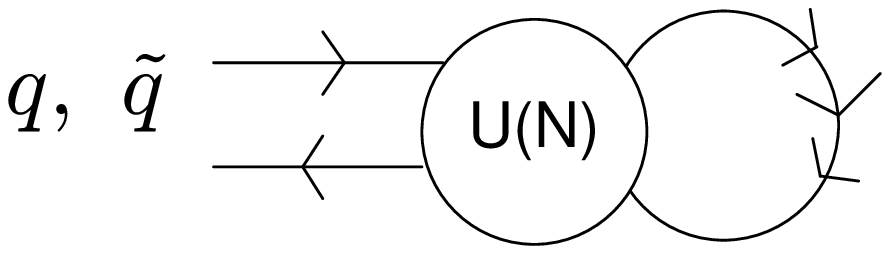} \end{array}$
 \caption{Quivers with
flavor for the D6 brane in $\IC\IP^3$ and for $V^{5,2}$.}
\end{figure}

With many D6 branes, the resulting M-theory geometry has a
singular horizon, locally giving the expected $\IZ_{N_f}$ orbifold
singularity that carries the $SU(N_f)$ gauge fields of the AdS
dual. It is described by $t\tilde{t} = (a_1 b_1)^{N_f}$ in the
same projective space.

This geometry can also be obtained from a IIB construction.
Putting $N$ D3 branes on a circle, and intersecting with an NS5
and $(1,k)$5 brane at $\CN=3$ angles engineers the YM-CSM theory
that flows in the infrared to the $\CN=6$ theory. As explained in
\cite{GJ, Kirsch, WeiLi, Fujita} the addition of a D5 brane at
$\CN=3$ angles introduces a fundamental hypermultiplet. In our
case, we rotate the D5 brane such that it is parallel with the NS5
brane in two planes, and (at zero IIB axion) perpendicular in the
other plane. This preserves $\CN=2$ supersymmetry. Starting
instead with the D5 brane at general $\CN=2$ preserving angles, it
will bend in the IR region, corresponding to the fact that the
coefficients in the superpotential $\CW = c_1 q A_1 B_1 \tilde q +
c_2 q A_2 B_2 \tilde q + m (q \tilde q)^2$ run under the
renormalization group flow. Generically the theory will flow to
the $\CN=3$ point, however when $c_2 = 0$, that term will not be
generated, and one will obtain the model discussed above. The
quartic term, $m (q\tilde q)^2$ appears to be absent in the theory
with $c_2= 0$, and likely $m$ would flow to 0 in that case.


\subsection{A non-toric example, $V^{5,2}$}

The cone over this Sasaki-Einstein manifold is a non-toric
Calabi-Yau 4-fold, $X^{5,2}$, is described by the hypersurface
$z_1^2 + ... + z_5^2 = 0$, with an obvious $SO(5) \times U(1)_R$
isometry group\footnote{I am grateful to I. Klebanov and S. Pufu
for very useful discussions on this issue.}. Picking out a $U(1)_B
\subset SO(5)$ isometry, the theory of $N$ M2 branes on
$X^{5,2}/\IZ_k$, for $\IZ_k \subset U(1)_B$, will have weakly
coupled gauge groups for large $k$. Such $U(1)$'s are specified by
two integers, however only two possibilities result in theories
with non-singular horizons (and possible D6 branes) in the IIA
reduction.

The first is a rotation in the $z_1$ - $z_2$ plane. This has no
orbifold fixed loci, but does have an ordinary fixed surface, $z_1
= z_2 = 0$. Therefore the reduction to IIA on this circle with
have an explicit D6 brane, and the corresponding 2+1 field theory
will have fundamental flavors.

The associated Calabi-Yau 3-fold is given by $X^{5,2}//U(1)_B =
\IC^3$, parameterized by $z_3, z_4, z_5$. The D6 branes are
wrapping the locus in the 7d cone $X^{5,2}/U(1)_B$ given by
$z_3^2+z_4^2 +z_5^2= 0$. Therefore we have a $U(N)$ pure
Yang-Mills theory, with 3 adjoint chirals, and $k$
fundamental/anti-fundamental pair, with superpotential $$W = \Tr
X[Y,Z] + q (X^2 + Y^2 + Z^2) \tilde q.$$ Although there is no
Chern-Simons term, even on the moduli space in this completely
non-chiral quiver, the gauge field becomes weakly interacting when
the number of flavors is large.

The other possibility is a simultaneous rotation in the $z_1$ -
$z_2$ and $z_4$ - $z_5$ planes. This has the fixed locus $z_1 =
z_2 = z_3 = z_4 = 0$ in $\IC^5$, but the defining equation then
implies that $z_5= 0$ as well, so it is an isolated singularity.
The resulting Lagrangian description of the dual to $V^{5,2}$ was
found in \cite{martelli-sparks}, and is based on the quiver
associated to D3 branes on the Calabi-Yau 3-fold $Y =
X^{5,2}//U(1)_B$, which can be rewritten as the equation $a_1 b_1
+ a_2 b_2 + z_5^2 = 0$ in $\IC \times $conifold, where $a_1 = z_1
+ i z_2$, $a_2 = z_3 + i z_4$, $b_1 = z_1 - i z_2$, and $b_2 = z_3
- i z_4$.

\subsection{Chiral flavors and $Q^{111}$}

Next I consider two simple examples with chiral flavors. First add
to the $\CN=6$ CSM theory a flavor $q$ entering one node and
$\tilde{Q}$ exiting the other, with a new term in the
superpotential, $\CW = q A_1 \tilde Q$. The Chern-Simons levels,
$k + \frac{1}{2}$, must be equal half integers with opposite sign.

 Then the usual monopole and anti-monopole
will pick up an additional gauge charge of $\frac{1}{2}$ under the
first $U(1)$ and $-\frac{1}{2}$ under the second at 1-loop. The
OPE will be given by $T \tilde T \sim A_1$, consistent with the
charges $(k+1) + (-k) = 1$ and the dimensions computed in section
3.

The variable $A_1$ can be eliminated from the description of the
chiral ring, resulting in the Calabi-Yau 4-fold $\IC^5//U(1)$,
where the group acts by $(a_2, b_1, b_2, t, \tilde t) \mapsto
(\lambda a_2, \lambda^{-1} b_1, \lambda^{-1} b_2, \lambda^{k+1} t,
\lambda^{-k} \tilde t)$. In particular $\IC \times$conifold is one
possibility when $k=0, -1$.

Adding $N_f$ fundamentals of this type changes the relation to
$t\tilde t = a_1^{N_f}$, and it is easy to check that scaling $k$
and $N_f$ together gives a simple quotient of the Calabi-Yau
4-fold along the direction of the $U(1)_B$ isometry, as expected.

The final example will involve a pair of such chiral fields, and
$\CW = q_1 A_1 \tilde Q_1 + q_2 A_2 \tilde Q_2$. Then the monopole
operators will have charges $k+1$ and $1-k$, where $\pm k$ are the
CS levels, and the OPE $T \tilde T \sim A_1 A_2$ is consistent
with gauge invariance and R-charge conservation.

This gives precisely the cone over $Q^{111}$ in the case that the
bare levels are vanishing! This configuration can be engineering
in IIB using an NS5 brane and a web of NS5 and D5 merging to make
a (1,1)5 brane. Therefore there should be a renormalization group
flow from the theory describing a pair of NS5 branes and one D5
brane to this $Q^{111}$ dual theory, after turning on an axial
mass to form the NS5, D5, $(1,1)$ 5 brane web. It would be
interesting to find the dual to that $\CN=2$ flow, which cannot
occur in four dimensional gauge theory. The fact that such a
fivebrane web results in an effective shift of the Chern-Simons
level when a chiral fundamental gets a real mass term on the
moduli space was used in \cite{BHKK} to derive the Chern-Simons
terms present on D3 branes stretched between $(1,k_i)$ fivebranes.

Further discussion of the IIB brane constructions, and the
detailed analysis and matching of the Kaluza-Klein harmonics on
the Sasaki-Einstein manifolds resulting from Chern-Simons-matter
theories with flavors will be investigated in the companion paper
\cite{JKP}.

Comparing these quivers with those found in \cite{Hanany} with the
same moduli space, one sees that the $U(1)_B$ actions on the
moduli spaces are identical. The latter fact means these CSM
theories cannot be strong-weak duals of each other - they are
weakly coupled in the same regime. It would be interesting to see
whether they could be related by the three dimensional version of
Seiberg duality \cite{giveon-kutasov} that generalizes fivebrane
moves to changes of basis in the quiver. Application of the rules
of \cite{BD} to the nodes with only a single arrow entering and
exiting and ungauging the dual nodes actually results in the same
quivers, now with flavors, that I have found. But it remains
unclear how the nodes would become ungauged; moreover from the
point of view of this paper, it is extremely natural to have
fundamental flavors given that the $U(1)_B$ isometry corresponding
to the M-circle has fixed points, resulting in D6 branes in the
IIA description dual to the gauge theory in the 't Hooft limit. If
such a duality were possible, the theories might describe
different values of the discrete torsion flux in M-theory.

\subsection*{Acknowledgments}
I would like to thank O.~Bergman, D.~Gaiotto and A.~Kapustin for
helpful discussions, F.~Benini and S.~Cremonesi for correspondence
prior to publication, and especially I.~Klebanov and S.~Pufu for
collaboration during the initial stages of this project. I am
supported in part by DOE grant DE-FG02-96ER40959.


\begin{thebibliography}{}

\bibitem{witten}
  E.~Witten,
  ``String theory dynamics in various dimensions,''
  Nucl.\ Phys.\  B {\bf 443}, 85 (1995)
  [arXiv:hep-th/9503124].

\bibitem{ABJM}
  O.~Aharony, O.~Bergman, D.~L.~Jafferis and J.~Maldacena,
  ``N=6 superconformal Chern-Simons-matter theories, M2-branes and their gravity duals,''
  arXiv:0806.1218 [hep-th].


\bibitem{BL}
  J.~Bagger and N.~Lambert,
  ``Modeling multiple M2's,''
  Phys.\ Rev.\  D {\bf 75}, 045020 (2007)
  [arXiv:hep-th/0611108].

\bibitem{BL2}
  J.~Bagger and N.~Lambert,
  ``Gauge Symmetry and Supersymmetry of Multiple M2-Branes,''
  Phys.\ Rev.\  D {\bf 77}, 065008 (2008)
  [arXiv:0711.0955 [hep-th]].

\bibitem{BL3}
  J.~Bagger and N.~Lambert,
  ``Comments On Multiple M2-branes,''
  JHEP {\bf 0802}, 105 (2008)
  [arXiv:0712.3738 [hep-th]].

\bibitem{G}
  A.~Gustavsson,
  ``Algebraic structures on parallel M2-branes,''
  Nucl.\ Phys.\  B {\bf 811}, 66 (2009)
  [arXiv:0709.1260 [hep-th]].

\bibitem{Schwarz}
  J.~H.~Schwarz,
  ``Superconformal Chern-Simons theories,''
  JHEP {\bf 0411}, 078 (2004)
  [arXiv:hep-th/0411077].

\bibitem{GY}
  D.~Gaiotto and X.~Yin,
  ``Notes on superconformal Chern-Simons-matter theories,''
  JHEP {\bf 0708}, 056 (2007)
  [arXiv:0704.3740 [hep-th]].

\bibitem{BKK}
  M.~K.~Benna, I.~R.~Klebanov and T.~Klose,
  ``Charges of Monopole Operators in Chern-Simons Yang-Mills Theory,''
  arXiv:0906.3008 [hep-th].


\bibitem{BCC} F.~Benini, C.~Closset and S.~Cremonesi, ``Chiral
flavors and M2-branes at toric CY4 singularities,''
arXiv:0911.4127 [hep-th].




\bibitem{GT}
  D.~Gaiotto and A.~Tomasiello,
  ``The gauge dual of Romans mass,''
  arXiv:0901.0969 [hep-th].

\bibitem{JT}
  D.~L.~Jafferis and A.~Tomasiello,
  ``A simple class of N=3 gauge/gravity duals,''
  JHEP {\bf 0810}, 101 (2008)
  [arXiv:0808.0864 [hep-th]].

\bibitem{MS}
  D.~Martelli and J.~Sparks,
  ``Moduli spaces of Chern-Simons quiver gauge theories and AdS(4)/CFT(3),''
  Phys.\ Rev.\  D {\bf 78}, 126005 (2008)
  [arXiv:0808.0912 [hep-th]].

\bibitem{HZ}
  A.~Hanany and A.~Zaffaroni,
  ``Tilings, Chern-Simons Theories and M2 Branes,''
  JHEP {\bf 0810}, 111 (2008)
  [arXiv:0808.1244 [hep-th]].



\bibitem{aganagic}
  M.~Aganagic,
  ``A Stringy Origin of M2 Brane Chern-Simons Theories,''
  arXiv:0905.3415 [hep-th].

\bibitem{GGPT}
  J.~P.~Gauntlett, G.~W.~Gibbons, G.~Papadopoulos and P.~K.~Townsend,
  ``Hyper-Kaehler manifolds and multiply intersecting branes,''
  Nucl.\ Phys.\  B {\bf 500}, 133 (1997)
  [arXiv:hep-th/9702202].

\bibitem{GW}
  D.~Gaiotto and E.~Witten,
  ``S-Duality of Boundary Conditions In N=4 Super Yang-Mills Theory,''
  arXiv:0807.3720 [hep-th].

\bibitem{KOO}
  T.~Kitao, K.~Ohta and N.~Ohta,
  ``Three-dimensional gauge dynamics from brane configurations with (p,q)-fivebrane,''
  Nucl.\ Phys.\  B {\bf 539}, 79 (1999)
  [arXiv:hep-th/9808111].

\bibitem{BHKK}
  O.~Bergman, A.~Hanany, A.~Karch and B.~Kol,
  ``Branes and supersymmetry breaking in 3D gauge theories,''
  JHEP {\bf 9910}, 036 (1999)
  [arXiv:hep-th/9908075].

\bibitem{martelli-sparks}
  D.~Martelli and J.~Sparks,
  ``AdS\_4/CFT\_3 duals from M2-branes at hypersurface singularities and their
  deformations,''
  arXiv:0909.2036 [hep-th].

\bibitem{ABJ}
  O.~Aharony, O.~Bergman and D.~L.~Jafferis,
  ``Fractional M2-branes,''
  arXiv:0807.4924 [hep-th].


\bibitem{Aharonyetal}
  O.~Aharony, A.~Hashimoto, S.~Hirano and P.~Ouyang,
  ``D-brane Charges in Gravitational Duals of 2+1 Dimensional Gauge Theories
  and Duality Cascades,''
  arXiv:0906.2390 [hep-th].

\bibitem{Kapustin1}
  V.~Borokhov, A.~Kapustin and X.~k.~Wu,
  ``Topological disorder operators in three-dimensional conformal field theory,''
  JHEP {\bf 0211}, 049 (2002)
  [arXiv:hep-th/0206054].

\bibitem{Kapustin2}
  V.~Borokhov, A.~Kapustin and X.~k.~Wu,
  ``Monopole operators and mirror symmetry in three dimensions,''
  JHEP {\bf 0212}, 044 (2002)
  [arXiv:hep-th/0207074].

\bibitem{Borokhov}
  V.~Borokhov,
  ``Monopole operators in three-dimensional N = 4 SYM and mirror symmetry,''
  JHEP {\bf 0403}, 008 (2004)
  [arXiv:hep-th/0310254].


\bibitem{jafferis-yin}
  D.~L.~Jafferis and X.~Yin,
  ``Chern-Simons-Matter Theory and Mirror Symmetry,''
  arXiv:0810.1243 [hep-th].


\bibitem{GJ}
  D.~Gaiotto and D.~L.~Jafferis,
  ``Notes on adding D6 branes wrapping RP3 in AdS4 x CP3,''
  arXiv:0903.2175 [hep-th].


\bibitem{Kirsch}
  S.~Hohenegger and I.~Kirsch,
  ``A note on the holography of Chern-Simons matter theories with flavour,''
  arXiv:0903.1730 [hep-th].

\bibitem{WeiLi}
  Y.~Hikida, W.~Li and T.~Takayanagi,
  ``ABJM with Flavors and FQHE,''
  JHEP {\bf 0907}, 065 (2009)
  [arXiv:0903.2194 [hep-th]].

\bibitem{Fujita}
  M.~Fujita and T.~S.~Tai,
  ``Eschenburg space as gravity dual of flavored N=4 Chern-Simons-matter
  theory,''
  JHEP {\bf 0909}, 062 (2009)
  [arXiv:0906.0253 [hep-th]].

\bibitem{CP}
  B.~Chandrasekhar and B.~Panda,
  ``Brane Embeddings in AdS\_4 x CP\^3,''
  arXiv:0909.3061 [hep-th].

\bibitem{AEMOW}
  M.~Ammon, J.~Erdmenger, R.~Meyer, A.~O'Bannon and T.~Wrase,
  ``Adding Flavor to AdS4/CFT3,''
  arXiv:0909.3845 [hep-th].


\bibitem{ouyang}
  P.~Ouyang,
  ``Holomorphic D7-branes and flavored N = 1 gauge theories,''
  Nucl.\ Phys.\  B {\bf 699}, 207 (2004)
  [arXiv:hep-th/0311084].

\bibitem{JKP}
 D.~Jafferis, I.~Klebanov and S.~Pufu,
 work in progress.

\bibitem{Hanany}
  S.~Franco, A.~Hanany, J.~Park and D.~Rodriguez-Gomez,
  ``Towards M2-brane Theories for Generic Toric Singularities,''
  JHEP {\bf 0812}, 110 (2008)
  [arXiv:0809.3237 [hep-th]].

\bibitem{giveon-kutasov}
  A.~Giveon and D.~Kutasov,
  ``Seiberg Duality in Chern-Simons Theory,''
  Nucl.\ Phys.\  B {\bf 812}, 1 (2009)
  [arXiv:0808.0360 [hep-th]].

\bibitem{BD}
  D.~Berenstein and M.~R.~Douglas,
  ``Seiberg duality for quiver gauge theories,''
  arXiv:hep-th/0207027.












\end{thebibliography}
\end{document}